\documentclass[aps,prl,showpacs,twocolumn,amsmath,amssymb,superscriptaddress]{revtex4}
\usepackage{graphicx}
\usepackage{Jonasmacros}
\usepackage{bm}
\usepackage{fancyheadings}
\begin{document}
\date{\today}
\title{Anomalous Conductance Suppression in Magnetic Clusters}
\author{J. Fransson}
\email{jonasf@lanl.gov}
\affiliation{Theoretical Division, Los Alamos National Laboratory, Los Alamos, New Mexico 87545, USA}
\affiliation{Center for Nonlinear Science, Los Alamos National Laboratory, Los Alamos, New Mexico 87545, USA}
\author{A. V. Balatsky}
\email{avb@lanl.gov}
\affiliation{Theoretical Division, Los Alamos National Laboratory, Los Alamos, New Mexico 87545, USA}
\affiliation{Center for Integrated Nanotechnology, Los Alamos National Laboratory, Los Alamos, New Mexico 87545, USA}

\begin{abstract}
Recent scanning tunneling microscope measurements on magnetic clusters reveal an anomalous conductance dip at zero bias voltage, whenever the cluster consists of an odd number of magnetic atoms [Science, {\bf 312}, 1021 (2006)]. We address the conductance anomaly within a model, and reproduce the conductance dip with the only assumption that the magnetic cluster has a conducting level near the Fermi level of the system. We show that the width of the conductance dip scales quadratically with external magnetic fields, in excellent agreement with experiments. In addition, we predict that the presence of the impurity will be measurable as inelastic Friedel oscillation in the substrate density of electron states.
\end{abstract}
\pacs{72.25.-b, 73.23.-b, 73.63.Kv}
\maketitle

The study of spin-dependent transport phenomena in mesoscopic systems is becoming a well established research field \cite{zutic2004} which has gained lots of attention recently, both experimentally \cite{tsukagoshi1999} and theoretically \cite{rudzinski2001}. While one major reason is the prospect of using such systems in spin-qubit readout \cite{bandyopadhyay2003}, fundamental questions concerning how spin coherence and interactions influence the transport are of main interest. Measurements of the effects can be performed by, for example, electron spin resonance (ESR) \cite{engel2001}, and more recently, ESR-scanning tunneling microscopy (ESR-STM) techniques \cite{manassen1989,durkan2002}. In this context, the ESR-STM is especially interesting because of the possibility to manipulate single spins \cite{manoharan2002,balatsky2002,koppens2006}, something which is crucial in spintronics and quantum information. Experimentally, modulation in the tunneling current has been observed by STM using spin-unpolarized electron beam \cite{engel2001,manassen1989}. Lately, there has also been a growing interest in using spin-polarized electron beam for direct detection of spin structures \cite{wiesendanger2000}, as well as utilizing the inelastic electron scanning tunneling spectroscopy (IETS) for detection of local spatial variations in electron-boson coupling in molecular systems \cite{madhaven2001,grobis2005}. IETS was recently suggested to be used for surface imaging of inelastic Friedel oscillations caused by inelastic scattering effects on a local impurity \cite{fransson(friedel)2007}.

Recent STM measurements of magnetic atoms, e.g. Mn, located on a metallic surface revealed conductance dips at vanishing bias voltage whenever the magnetic cluster consisted of an odd number of atoms \cite{heinrich2004,hirjibehedin2006}. The conductance dips were assigned to inelastic scattering, e.g. spin-flip, events between the spin channels in the system. In case of an even number of magnetic atoms in the cluster the zero bias dip were absent, however, the conductance were suppressed for a larger range of biases between. The limits of this range were determined by the energy barrier needed for spin-flip transitions between the singlet $(S=0,\ m_z=0)$ ground state and the  triplet states $(S=1,\ m_z=0,\pm1)$.

\begin{figure}[b]
\begin{center}
\includegraphics[width=8.5cm]{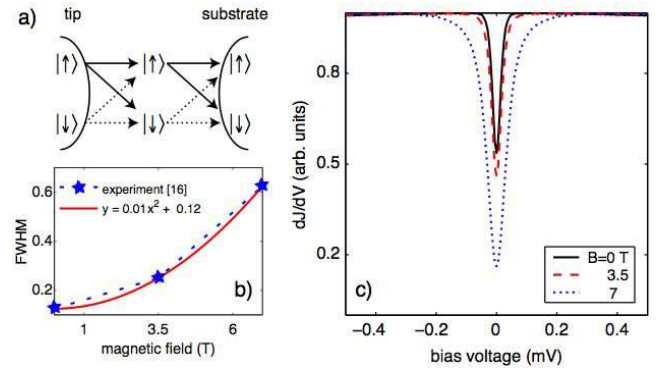}
\end{center}
\caption{(Color online) a) Sketch of the branching of the tunneling electron wave between the two spin-channels due to the strong spin-flip tunneling rate. b) Quadratic fit of the FWHM of the experimental conductance dips, extracted from Ref. \cite{hirjibehedin2006}, as function of the external magnetic field. c) Calculated differential conductance for different external magnetic fields within the present theory, e.g. using Eq. (\ref{eq-T}) with equal spin-preserving ($\Gamma_0$) and spin-flip ($\Gamma_s$) tunneling rates ($\Gamma_s=\Gamma_0$).}
\label{fig-djdv}
\end{figure}

The central question we address in this paper is the anomalous feature at zero bias voltage seen in the STM conductance for an odd number of magnetic atoms in the cluster. In this case the ground state of the magnetic cluster, which is anti-ferromagnetically ordered \cite{bergman2006}, is a doubly degenerate spin state. Due to (inelastic) spin-flip transitions in the tunneling between the magnetic cluster and the STM tip and the substrate, further channels open for conduction in addition to the existing spin-preserving channels. The result is that the tunneling electrons become shared between different pathways in the system, see Fig. \ref{fig-djdv} a), which gives rise to Fano-like interference effects between tunneling electron wavefunctions. The interference tends to suppress the conductance through one of the channels whereas the conductance in the other is enhanced. The destructive interference in the former case, generates a markedly suppressed conductance around zero biases whenever the spin-flip tunneling rate is of the order of the spin-preserving tunneling rate.

The result of this Letter is that we reproduce the conductance dip within a model with the only assumption that the magnetic cluster has a conducting level at the Fermi level of the system. Through the model, we find that the width of the conductance dip scales quadratically with external magnetic field $B$, in excellent agreement with experiments, c.f. Fig. \ref{fig-djdv} b).

We consider a model where electrons tunnel between the tip and the substrate via a molecular level $\dote{\sigma}$, assumed to be close to the common equilibrium chemical potential $\dote{F}$ of the tip | atomic cluster | substrate-system. We argue that we can easily reproduce the dip in the tunneling conductance for odd spin state within this model, with the only assumption that $\dote{\sigma}\lesssim\dote{F}$. For simplicity, we assume that there are no electron-electron or spin-flip interactions in the metallic leads. The spin-flip transitions are confined in both barriers between the local spin and the leads. The system is modeled by the following Hamiltonian: 
\begin{eqnarray}
\Hamil&=&\sum_{k\sigma}\leade{k}\cdagger{k}\cc{k}
	+\sum_\sigma\dote{\sigma}\ddagger{\sigma}\dc{\sigma}
\nonumber\\&&
	+\sum_{k\sigma\sigma'}(v_{k\sigma\sigma'}\cdagger{k}\dc{\sigma'}+H.c.),
\label{eq-model}
\end{eqnarray}
where $\cdagger{k}\ (\cc{k})$ creates (annihilates) an electron with momentum $k$, spin $\sigma=\up,\down$, and energy $\leade{k}$ in the left $(L)$ or right $(R)$ lead, whereas $\ddagger{\sigma}\ (\dc{\sigma})$ creates (annihilates) an electron with spin $\sigma$ at the energy $\dote{\sigma}=\dote{0}+\sigma\Delta/2$ on the local spin site, where $\Delta$ is the spin split on the impurity spin, which may be caused by external magnetic field $\bfB$ and/or the internal structure of the magnetic cluster. Electrons on the spin site hybridizes with those in the leads with the rate $v_{k\sigma\sigma'}$. Consider the case of no spin-orbit interaction but purely exchange interacting atomic orbitals with single electronic spin, e.g. $J\bfS_\text{atomic}\cdot\bfs_\text{electronic}$. The Fermi golden rule already would indicate that the lifetime broadening $\Gamma_\sigma$ of such electronics states as well as the spin-flip scattering rate $\Gamma_s$ all are proportional to $J^2[N_L(\dote{F})+N_R(\dote{F})]$, where $N_{L/R}(\dote{F})$ is the Fermi level density of states (DOS) in the tip/substrate.

The current through the system is derived by standard methods, considering the temporal derivative of the occupation $N_L=\sum_{p\sigma\in L}\cdagger{p}\cc{p}$ in the tip (left lead), i.e. $I=-e\ddtinline\av{N_L}$. The differential conductance ${\cal G}(V)=\partial I/\partial V$ can then be written as \cite{meir1992}
\begin{equation}
{\cal G}(V)=\frac{e^2\beta}{4h}\tr\int
	{\cal T}(\omega)\cosh^{-2}\beta(\omega-eV)d\omega,
\label{eq-G}
\end{equation}
where ${\cal T}(\omega)=\bfGamma^L\bfG^r(\omega)\bfGamma^R\bfG^a(\omega)$ is the transmission coefficient. Here, the coupling between the local spin and the tip/substrate is given by $\bfGamma^{L/R}$, with $\Gamma_{\sigma\sigma'}^{L/R}=2\pi\sum_{k\sigma''\in L/R}v_{k\sigma''\sigma}^*v_{k\sigma''\sigma'}\delta(\omega-\dote{k\sigma''})$, whereas $\bfG^{r/a}(\omega)$ is the Fourier transformed retarded/advanced Green function (GF) for electrons on the local spin site: $G_{\sigma\sigma'}^{r/a}(t)=\mp i\theta(\pm t)\av{\anticom{\dc{\sigma}(t)}{\ddagger{\sigma'}(0)}}$. Note that both $\Gamma^{L/R}$ and $G^{r/a}$ are matrices in the spin space of the impurity.

The impurity site GF is solved by equation of motion, which yields (defining $A_\sigma\equiv A_{\sigma\sigma}$ for shorter notation)
\begin{subequations}
\label{eq-GF}
\begin{eqnarray}
G_\up^r(\omega)&=&\frac{\omega-\dote{\down}-\Sigma_\down^r}
	{(\omega-\omega^r_+)(\omega-\omega^r_-)},
\label{eq-gfup}\\
G_{\up\down}^r(\omega)&=&\frac{\Sigma_{\up\down}^r}
	{(\omega-\omega^r_+)(\omega-\omega^r_-)},
\label{eq-gfud}
\end{eqnarray}
\end{subequations}
where the self-energy matrix due to the couplings to the leads is
\begin{equation}
\Sigma_{\sigma\sigma'}^r=\sum_{k\sigma''\in L\cup R}
	\frac{v_{k\sigma''\sigma}^*v_{k\sigma''\sigma'}}
	{\omega-\dote{k\sigma''}+i0^+}.
\label{eq-S}
\end{equation}
The GFs $G^r_\down$ and $G_{\down\up}^r$ are obtained by exchanging $\up\leftrightarrow\down$ in the above equations. For simplicity we assume large band-witdths in the leads, hence, we can approximate the self-energies by $\Sigma^r_{\sigma\sigma'}=-i\pi\sum_{k\sigma''}v_{k\sigma''\sigma}^*v_{k\sigma''\sigma'}\delta(\omega-\dote{k\sigma''})=(-i/2)(\Gamma^L_{\sigma\sigma'}+\Gamma^R_{\sigma\sigma'})$. In the present case, where the magnetic atom is located on a substrate, the coupling between the impurity site and the tip (left lead) is much smaller than the coupling between the substrate and the impurity site, i.e. $\Gamma^L_{\sigma\sigma'}/\Gamma^R_{\sigma\sigma'}=\lambda\ll1$. Hence, the self-energies for the impurity site electrons are then dominated by the coupling to the substrate (right lead), hence, $\Sigma^r_{\sigma\sigma'}\approx-i\Gamma^R_{\sigma\sigma'}/2$. Hermiticity of the couplings also require that $\Gamma^{L/R}_{\up\down}=\Gamma^{L/R}_{\down\up}=\Gamma^{L/R}_s$. In the experimental setup, the tunneling current is spin-independent \cite{heinrich2004,hirjibehedin2006}. Hence, it is reasonable to let $\Gamma^L_\sigma=\Gamma^L_0$ and $\Gamma^R_\sigma=\Gamma^R_0$, where $\Gamma^L_0=\lambda\Gamma_0$ and $\Gamma^R_0=\Gamma_0$. In this case, the poles $\omega_\pm^r$ of the GFs in Eq. (\ref{eq-GF}) are given by
\begin{equation}
\omega_\pm^r=\dote{0}-i(\Gamma_0
	\mp\sqrt{\Gamma_s^2-\Delta^2})/2.
\label{eq-poles}
\end{equation}

Suppose that the spin-split $\Delta=g\mu_BB$, where $g$ and $\mu_B$ is the gyromagnetic ratio and Bohr magneton, respectively. We assume that $\Delta\ll\Gamma_0,\Gamma_s$, which is reasonable for tunneling rates of the order 10 meV. In this case, the widths of the poles scale quadratically with the magnetic field, since the width can be written
\begin{equation}
\Gamma_0\mp\Gamma_s\sqrt{1-\biggl(\frac{\Delta}{\Gamma_s}\biggr)^2}
	\approx\Gamma_0\mp\Gamma_s\left[1-
		\frac{1}{2}\biggl(\frac{g\mu_BB}{\Gamma_s}\biggr)^2\right].
\label{eq-widths}
\end{equation}
As seen, in the absence of magnetic field $B$, the pole $\omega_+$ acquires a vanishing width as $\Gamma_s\rightarrow\Gamma_0$, while the width of the pole $\omega_-$ appoaches $\Gamma_0$, which is also clear from Eq. (\ref{eq-poles}). Physically this means that an electron in the level $\omega_+$ only weakly interacts with the conduction electrons in the leads. Therefore, the conduction through the magnetic cluster tends to decrease in the limit $\Gamma_s\rightarrow\Gamma_s$, since then the electron occupation of the level $\omega_+$ approaches unity whenever the localized level $\dote{0}\leq\dote{F}$..

Hence, the conditions that yield the vanishingly small width of the pole $\omega_+$ give rise to an anti-resonance in the transmission ${\cal T}(\omega)$ at $\omega=\dote{0}$. This is easily seen by substituting the expression for the GF $\bfG^{r/a}(\omega)$ on the local spin site into the expression for the transmission, which gives
\begin{eqnarray}
{\cal T}(\omega)&=&\frac{\lambda}{2}\frac{\Gamma_0^2+\Gamma_s^2}
	{|(\omega-\omega_+^r)(\omega-\omega_-^r)|^{2}}
\nonumber\\&\times&
	\biggl[4(\omega-\dote{0})^2
		+\frac{\Gamma_0^2-\Gamma_s^2}{\Gamma_0^2+\Gamma_s^2}
			(\Delta^2+\Gamma_0^2-\Gamma_s^2)\biggr].
\label{eq-T}
\end{eqnarray}
When the spin-flip tunneling rate is comparable to the spin-preserving tunneling rate, e.g. $\Gamma_s\approx\Gamma_0$, the last term in this expression vanishes, leading to the transmission
\begin{equation}
{\cal T}(\omega)\sim\frac{(\omega-\dote{0})^2}
	{(\omega-\dote{0})^4+(\omega-\dote{0}^2)(\Gamma_0^2-\Delta^2/2)+(\Delta/2)^4},
\label{eq-T1}
\end{equation}
which exactly vanishes for $\omega=\dote{0}$ and all finite magnetic fields. Thus, assuming that the local level $\dote{0}\simeq0$, the differential conductance of the systems will display a dip for vanishing bias voltages, see Fig. \ref{fig-djdv} c). Absence of magnetic fields leads to $\Delta=0$, which reduces Eq. (\ref{eq-T1}) to
\[
{\cal T}(\omega)\sim\frac{1}{(\omega-\dote{0})^2+\Gamma_0^2},
\]
that is, a single peak around $\dote{0}$ of width $\Gamma_0$. The width of the anti-resonance in the transmission is related to the width of the poles $\omega_+$. Thus, the width of the anti-resonance scales quadratically with the magnetic field according to Eq. (\ref{eq-widths}), something which is also seen in Fig. \ref{fig-djdv} c). This feature is in excellent agreement with the experimental findings in Ref. \cite{hirjibehedin2006}.

\begin{figure}[t]
\begin{center}
\includegraphics[width=8.5cm]{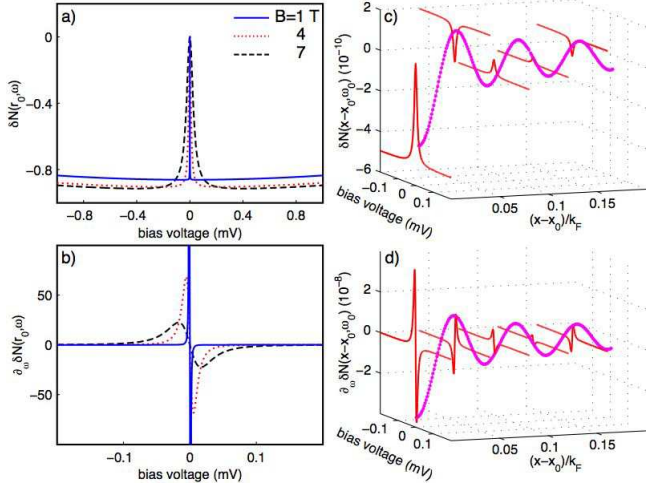}
\end{center}
\caption{(Color online) a) Correction $\delta N(\bfr_0,\omega)$ to the local substrate DOS, and b) frequency derivative $\partial_\omega\delta N(\bfr_0,\omega)$ at the impurity. Spatial and frequency dependence of c) the correction $\delta N(\bfr,\omega)$, and the derivative $\partial_\omega\delta N(\bfr,\omega)$ at $B=4$ T. The bold lines display the spatial Friedel oscillations that would be measured in the substrate DOS, whereas the thin lines show $\partial_\omega\delta N(\bfr_i,\omega)$ and $\partial_\omega\delta N(\bfr_i,\omega)$, that would be measured at position $\bfr_i$.}
\label{fig-surf}
\end{figure}

The dip in the transmission can be understood as an effect of destructive interference of the branched tunneling electron wave functions between the two spin channels in the system. This interference effect also becomes visible in the local DOS in the substrate caused by the presence of the localized magnetic atom. In the absence of the magnetic impurity the local DOS of the substrate is featureless. In terms of GFs, the  unperturbed DOS is given by $N_0^R\equiv-(1/\pi)\im\sum_{q\sigma}g_{q\sigma}^r(\omega)=\sum_{q\sigma}\delta(\omega-\leade{q})$, where $g_{q\sigma}^r(\omega)$ is the Fourier transform of the retarded substrate GF $g_{q\sigma}(t)=\eqgr{\cc{q}(t)}{\cdagger{q}(0)}$ is the retarded unperturbed GF of the substrate. The presence of an impurity on the substrate, will then manifest itself in the local substrate DOS as a mirror image of the local DOS of the impurity, since the substrate GF, now redefined as $g_{q\sigma q'\sigma'}(t)=\eqgr{\cc{q}(t)}{\csdagger{q'\sigma'}(0)}$, becomes dressed by the presence of the impurity. Accordingly, the Fourier transformed retarded substrate GF can be factored as
\begin{equation}
g_{q\sigma q'\sigma'}^r=
	\delta_{qq'}\delta_{\sigma\sigma'}g_{q\sigma}^r
	+g_{q\sigma}^rv_{q\sigma\sigma_1}G_{\sigma_1\sigma_2}^r
		v_{q'\sigma'\sigma_2}^*g_{q'\sigma'}^r,
\label{eq-gsub}
\end{equation}
where we assume summation over repeated indices. The total 
DOS is then given by $N^R(\omega)=N_0^R+\delta N^R(\omega)$, where
\begin{eqnarray}
\lefteqn{
\delta N^R(\omega)\approx\frac{1}{2}\tr\im\bfGamma^R\bfG^r(\omega)
}
\label{eq-deltaN}\\&=&
	-\frac{(\Gamma_0^2+\Gamma_s^2)(\omega-\dote{0})^2
		+\frac{1}{4}
		(\Gamma_0^2-\Gamma_s^2)(\Gamma_0^2-\Gamma_s^2+\Delta^2)}
	{2|(\omega-\omega_+)(\omega-\omega_-)|^2}.
\nonumber
\end{eqnarray}
Furthermore, in the limit $\Gamma_s=\Gamma_0$ we find that the frequency derivative $\partial_\omega\delta N^R(\omega)$ is given by
\begin{eqnarray}
\lefteqn{
\frac{\partial}{\partial\omega}\delta N^R(\omega)=
	-2\Gamma_0^2
}
\label{eq-dwdN}\\&&
	\times\frac{(\omega-\dote{0})[(\omega-\dote{0})^4+(\Delta/2)^4]}
		{[(\omega-\dote{0})^4+(\omega-\dote{0})^2(\Gamma_0^2-\Delta^2/2)
			+(\Delta/2)^4]^2}.
\nonumber
\end{eqnarray}
Clearly, the presence of the impurity provides a fingerprint in the local substrate DOS, such that the density of the impurity manifests itself as a dip around $\omega=\dote{0}$ in the substrate DOS as shown in Fig. \ref{fig-surf} a) and b). Additionally, the anomalous conductance dip will be identified as a peak in the substrate DOS exactly at $\omega=\dote{0}$, Fig. \ref{fig-surf} a). This is understood since an electron in the level $\omega_+$ interacts only weakly with the electrons in the substrate and will, hence, have a negligible influence on the substrate DOS. Therefore, the substrate DOS returns to its undisturbed value at $\omega=\dote{0}$.

The result for the substrate DOS indicates measurability of the conductance dip as inelastic Friedel oscillations in real space on the surface. In order to see this, consider the real space GF of the substrate using Eq. (\ref{eq-gsub})
\begin{equation}
g(\bfr,\bfr';\omega)=g_0(\bfr,\bfr';\omega)
	+g_0(\bfr,0;\omega)\Sigma(\omega)g_0(0,\bfr';\omega).
\label{eq-grs}
\end{equation}
In two dimensions, the bare GF becomes
\begin{eqnarray}
\lefteqn{
g_0(\bfr,\bfr';\omega)=\sum_{\bfq}g_0(\bfq,\omega)e^{i\bfq\cdot(\bfr-\bfr')}
}
\nonumber\\&\approx&
	2\pi\frac{m}{\hbar^2}J_0(k_F|\bfr-\bfr'|)
		\biggl(\log\left|\frac{\omega+D}{\omega-D}\right|-i\pi\biggr),
\label{eq-g0rs}
\end{eqnarray}
where $k_F$ is the Fermi wave vector and $J_0(x)$ is the zero Bessel function, whereas the self-energy is identified by $\Sigma(\omega)=(1/2)\tr\bfGamma^R\bfG^r(\omega)$. The correction to the local substrate DOS is given by
\begin{equation}
\delta N(\bfr,\omega)=-\frac{1}{\pi}\im\{g_0(\bfr,0;\omega)
	\Sigma(\omega)g_0(0,\bfr;\omega)\}.
\label{eq-dN}
\end{equation}
The effects from the inelastic scattering responsible for the conductance dip are included in the self-energy, thus we evaluate $\partial\delta N(\bfr,\omega)/\partial\omega$ which is directly proportional to $d^2I/dV^2$ in the transport measurements. The expected sharp features at $\dote{0}$ is connected to the spin-flip tunneling. Noting that $\re{\{g_0\}}\approx0$, our calculations simplify to $\delta N(\bfr,\omega)\approx-(1/\pi)g_0(\bfr,0;\omega)[\im\Sigma(\omega)]g_0(0,\bfr;\omega)$, which gives, c.f. Eq. (\ref{eq-deltaN}), 
\begin{equation}
\delta N(\bfr,\omega)\approx\frac{\pi^3}{2}\left(\frac{2m}{\hbar^2}\right)^2
	J_0^2(k_Fr)\tr\im\bfGamma^R\bfG^r(\omega),
\label{eq-dNexp}
\end{equation}
where $r=|\bfr|$, and a corresponding expression for the frequency derivative $\partial_\omega\delta N(\bfr,\omega)$. This shows that the inelastic (spin-flip) tunneling between the substrate and the magnetic impurity will be manifested as inelastic Friedel oscillations in the spatially resolved substrate DOS at the energy $\omega=\dote{0}$, see Fig. \ref{fig-surf} c) and d).

In summary, within a model we reproduce the anomalous conductance dips in the STM measurements of magnetic clusters with an odd number of magnetic atoms. We show that the conductance dip is caused by a local level in the magnetic cluster which has a unit occupation probability and which hybridizes only weakly with the conduction bands in the tip and substrate. The weak hybridization between the level and the conduction bands is generated by strong spin-flip tunneling rate. We demonstrate that the isolated local level will manifest its presence through inelastic Friedel oscillations in spatially resolved STM measurements of the substrate DOS.

This work has been supported by US DOE, LDRD and BES, and was carried out under the auspices of the NNSA of the US DOE at LANL under Contract No. DE-AC52-06NA25396.


\begin{thebibliography}{20}
\bibitem{zutic2004} I. Zutic, J. Fabian, and S. Das Sarma, Rev. Mod. Phys. {\bf 76}, 323 (2004).
\bibitem{tsukagoshi1999} K. Tsukagoshi, B. W. Alphenaar, and H. Ago, Nature {\bf 401}, 572 (1999);
D. Orgassa, G. J. Mankey, and H. Fujiwara, Nanotechnology {\bf 12}, 281 (2001);
M. M. Deshmukh and D. C. Ralph, Phys. Rev. Lett. {\bf 89}, 266803 (2002);
B. Zhao, I. M\"onch, H. Vinzelberg, T. M\"uhl, and C. M. Schneider, Appl. Phys. Lett. {\bf 80}, 3144 (2002);
J. -R. Kim, H. M. So, J. -J. Kim, and J. Kim, Phys. Rev. B, {\bf 66}, 233401 (2002);
J. R. Petta, S. K. Slater, and D. C. Ralph, Phys. Rev. Lett, {\bf 93}, 136601 (2004).
\bibitem{rudzinski2001} W. Rudzinski and J. Barnas, Phys. Rev. B, {\bf 64}, 085318 (2001);
J. Fransson, O. Eriksson, and I. Sandalov, Phys. Rev. Lett. {\bf 88}, 226601 (2002);
J. K\"onig and J. Martinek, Phys. Rev. Lett. {\bf 90}, 166602 (2003);
M. -S. Choi, D. S\'anchez, and R. L\'opez, Phys. Rev. Lett. {\bf 92}, 056601 (2004);
J. Fransson, Europhys. Lett. {\bf 70}, 796 (2005).
\bibitem{bandyopadhyay2003} S. Bandyopadhyay, Phys. Rev. B, {\bf 67}, 193304 (2003).
\bibitem{koehler1993} J. K\"ohler, J. A. J. M. Disselhorst, M. C. J. M. Donckers, E. J. J. Groenen, J. Schmidt, and W. E. Moerner, Nature {\bf 363}, 242 (1993); J. Wrachtrup, C. von Borczyskowski, J. Bernard, M. Orrit, and R. Brown, \emph{ibid}. {\bf 363}, 244 (1993); Phys. Rev. Lett. {\bf 71}, 3565 (1993).
\bibitem{engel2001} H. -A. Engel and D. Loss, Phys. Rev. Lett. {\bf 86}, 4648 (2001); Phys. Rev. B, {\bf 65}, 195321 (2002).
\bibitem{manassen1989} Y. Manassen, R. J. Hamers, J. E. Demulth, and A. J. Castellano, Jr., Phys. Rev. Lett. {\bf 62}, 2531 (1989); D. Shachal and Y. Manassen, Phys. Rev. B, {\bf 46}, 4795 (1992); Y. Manassen, J. Magn. Reson. {\bf 126}, 133, (1997); Y. Manassen, I. Mukhopadhyay, and N. Ramesh Rao, Phys. Rev. B, {\bf 61}, 16223 (2000).
\bibitem{durkan2002} C. Durkan and M. E. Welland, Apply. Phys. Lett. {\bf 80}, 458 (2002).
\bibitem{manoharan2002} H. Manoharan, Nature {\bf 416}, 24 (2002); H. Manoharan, C. P. Lutz, and D. Eigler, Nature {\bf 403}, 512 (2000).
\bibitem{balatsky2002} A. V. Balatsky and I. Martin, Quantum Inf. Process. {\bf 1}, 53 (2002); A. V. Balatsky, Y. Manassen, and R. Salem, Phys. Rev. B, {\bf 66}, 195416 (2002).
\bibitem{koppens2006} F. H. L. Koppens, C. Buizert, K. J. Tielrooij, I. T. Vink, K. C. Nowack, T. Meunier, L. P. Kouwenhoven, and L. M. K. Vandersypen, Nature {\bf 442}, 766 (2006).
\bibitem{wiesendanger2000} S. Heinze, . Bode, A. Kubetzka, O. Pietzsch, X. Nie, S. Bl\"ugel, and R. Wiesendanger, Science {\bf 288}, 1805 (2000); A. Kubetzka , M. Bode, O. Pietzsch, and R. Wiesendanger, Phys. Rev. Lett. {\bf 88}, 057201 (2002); A. Wachowiak, J. Wiebe, M. Bode, O. Pietzsch, M. Morgenstern, and R. Wiesendanger, Science {\bf 298}, 577 (2002); J. Wiebe, A. Wachowiak, F. Meier, D. Haude, T. Foster, M. Morgenstern, and R. Wiesendanger, Rev. Sci. Instrum. {\bf 75}, 4871 (2004).
\bibitem{madhaven2001} V. Madhaven, W. Chen, T. Jamneala, M. F. Crommie, and N. S. Wingreen, Phys. Rev. B, {\bf 64}, 165412 (2001).
\bibitem{grobis2005} M. Grobis, K. H. Khoo, R. Yamachika, X. Lu, K. Nagaoka, S. G. Louie, M. F. Crommie, H. Kato, and H. Shinohara, Phys. Rev. Lett. {\bf 94}, 136802 (2005).
\bibitem{fransson(friedel)2007} J. Fransson and A. V. Balatsky, unpublished (2007); cond-mat/0701606.
\bibitem{heinrich2004} A. J. Heinrich, J. A. Gupta, C. P. Lutz, and D. M. Eigler, Science, {\bf 306}, 466 (2004).
\bibitem{hirjibehedin2006} C. F. Hirjibehedin, C. P. Lutz, and A. J. Hienrich, Science, {\bf 312}, 1021 (2006).
\bibitem{bergman2006} A. Bergman, L. Nordstr\"om, A. B. Klautau, S. Frota-Pess\^oa, and O. Eriksson, Phys. Rev. B, {\bf 73}, 174434 (2006).
\bibitem{meir1992} Y. Meir and N. S. Wingreen, Phys. Rev. Lett. {\bf 68}, 2512 (1992).

\end{thebibliography}
\end{document}